\documentclass[aps,pre,showpacs,twocolumn,superscriptaddress,floatfix]{revtex4}
\usepackage{graphicx}

\begin{document}
\title{Spatio-temporal patterns in the Hantavirus infection}

\date{\today}

\author{G. Abramson}
\email{abramson@cab.cnea.gov.ar}
\affiliation{Center for Advanced Studies and Department of Physics
and Astronomy, University of New Mexico, Albuquerque, New Mexico 87131}
\affiliation{Centro At\'{o}mico Bariloche and CONICET, 8400 S. C. de Bariloche, Argentina}

\author{V. M. Kenkre}
\email{kenkre@unm.edu}
\affiliation{Center for Advanced Studies and Department of Physics
and Astronomy, University of New Mexico, Albuquerque, New Mexico 87131}

\begin{abstract}
We present a model of the infection of Hantavirus in deer mouse,
\emph{Peromyscus maniculatus}, based on biological observations of the system
in the North American Southwest. The results of the analysis shed light on
relevant observations of the biological system, such as the sporadical
disappearance of the infection, and the existence of foci or ``refugia'' that
perform as reservoirs of the virus when environmental conditions are less than
optimal.
\end{abstract}


\maketitle

\section{Introduction}

Hantaviruses are infectious agents carried by rodents throughout the whole
world~\cite{schmaljohn97,mills99a,mills99b}. Some of them are able to cause
severe disease in humans, with a mortality rate of around 50\%, as in the case
of the Hantavirus Pulmonary Syndrome (HPS) caused by the Sin Nombre Virus in
the North American Southwest, or the Andes Virus in Patagonia. With few
exceptions, each hantavirus is associated with a single primary rodent host
species, chronically infected, and infects humans that come into contact with
it or its excreta. Sin Nombre Virus is primarily carried by the deer mouse,
\emph{Peromyscus maniculatus}, the most numerous mammal in North America,
prevalent throughout the region. It was the cause of an outbreak of fatal
pulmonary disease in the Southwest of the United States in 1993, when the virus
was first isolated and described. Since then, a great effort has been devoted
to understand the nature of the virus reservoir, its temporal and spatial
dynamics, and its relation to the human population, in an attempt to ultimately
identify and predict the risk of the disease.

Needless to say, a complete mathematical description of the dynamics of the
biological system, comprising the virus, the mice, the humans and the
environment, is a daunting task. The goal of the present investigation is much
less ambitious. From the biological complexities we  extract a few major
components centered on the basic ecological and epidemiological features of the
mice population. As the motivation for our analysis we choose two observed
characteristics of the disease. Both arise from  the fact that environmental
conditions strongly affect the dynamics and persistence of the infection. One
of them, a temporal characteristic, is the reported observation that the
infection can completely disappear from a population of mice if environmental
conditions are inadequate, only to reappear sporadically or when conditions
change~\cite{mills99b,calisher99,parmenter99}. The other, a spatial
characteristic, is that there are indications of ``focality'' of the infection
in ``reservoir'' populations~\cite{mills99b,kuenzi99}; as environmental changes
occur, these ``refugia''~\cite{yates2001} of the reservoir can expand or
contract, carrying the infection to other places.

The model we introduce incorporates the decay by death of the mice population,
the spread of the infection through their interaction, the increase by birth
and effect of the environment to stabilize the population, and also their
movement as a process of diffusion. We begin in Section~\ref{basic} by first
omitting the last feature (the movement from one location to another),
motivating the different dynamical mechanisms, and obtaining some basic results
including the observed temporal behavior. We proceed in Section~\ref{extended}
to perform a spatial extension of  the model to include movement and obtain
results relating to the refugia. A summary is given in the final section.

\section{Basic model of mouse population}

\label{basic}

We can incorporate the basic ingredients of the biological system in a model of
the mouse population only~\cite{anderson}. We suppose that the whole population
is composed of two classes of mice, susceptible and infected, represented by
$M_S$ and $M_I$ respectively. Sex and age composition of the population are
disregarded in this basic model. The temporal evolution of $M_S$ and $M_I$
contains two basic ingredients: the contagion of the infection, that converts
susceptible into infected, and a population dynamics independent of the
infection:
\begin{eqnarray}
\frac{dM_S}{dt}&=&b\,M -c M_S -\frac{M_S M}{K} -a\,M_S M_I, \label{dmsdt}\\
\frac{dM_I}{dt}&=&-c\,M_I -\frac{M_I M}{K} +a\,M_S M_I, \label{dmidt}
\end{eqnarray}
where $M_S$ and $M_I$ are the populations (or densities) of susceptible and
infected mice, respectively, and $M(t)=M_S(t)+M_I(t)$ is the total population
of mice. The motivation for the terms in Eqs.~(\ref{dmsdt},\ref{dmidt})
follows.

\emph{Births:} $b\,M$ represents births of mice, all of them born susceptible,
at a rate proportional to the total density, since all mice contribute equally
to the procreation~\cite{mills99b}.

\emph{Deaths:} $c$ represents the rate of depletion by death for natural reasons,
proportional to the corresponding density. If necessary, separate rates $c_S$
and $c_I$ could be introduced for the susceptible and infected populations
respectively.

\emph{Competition:} $-M_{S,I} M/K$ represent a limitation process in the population
growth, due to competition for shared resources. Each is proportional to the
probability of an encounter of a pair formed by one mouse of the corresponding
class, susceptible or infected, and one mouse of any class (since every mouse,
either susceptible or infected, has to compete with the whole population). $K$
is a ``carrying capacity,'' characterizing in a simplified way the capacity of
the medium to maintain a population of mice. Higher values of carrying capacity
represent a higher availability of water, food, shelter and other resources
that mice can use to thrive~\cite{murray}.

\emph{Infection:} $aM_IM_S$ represents the number of susceptible mice that
get infected, due to an encounter with an infected (and consequently
infectious) mouse, at a rate $a$ that we assume constant. More elaborate models
could incorporate a density dependence on $a$, for example due to an increased
frequency of fights, during which contagion occurs through bites, when the
density is too high and the population feels overcrowded~\cite{calisher99}. The
infection is chronic, infected mice do not die of it, and infected mice do not
lose there infectiousness probably for their whole life
\cite{mills99b,kuenzi99}. For these reasons, this single term adequately
describes the infection dynamics of the two subpopulations.

The sum of the two equations~(\ref{dmsdt},\ref{dmidt}) reduces to a single
equation for the whole population of logistic form:
\begin{equation}
\frac{dM}{dt}=(b-c)M\left(1-\frac{M}{(b-c)\,K}\right).
\label{dmdt}
\end{equation}
Logistic growth has been observed in laboratory populations of
\emph{Peromyscus}~\cite{terman68}, and is a well established metaphor of the dynamics
of a self limitating population~\cite{murray}.

There are four parameters that characterize the system
(\ref{dmsdt},\ref{dmidt}), viz. $a$, $b$, $c$ and $K$. Of these, we will choose
$K$ as a control parameter of the dynamics, since it is the one that best
represents the influence of the environment.

\begin{figure}
\centering
\resizebox{\columnwidth}{!}{\includegraphics{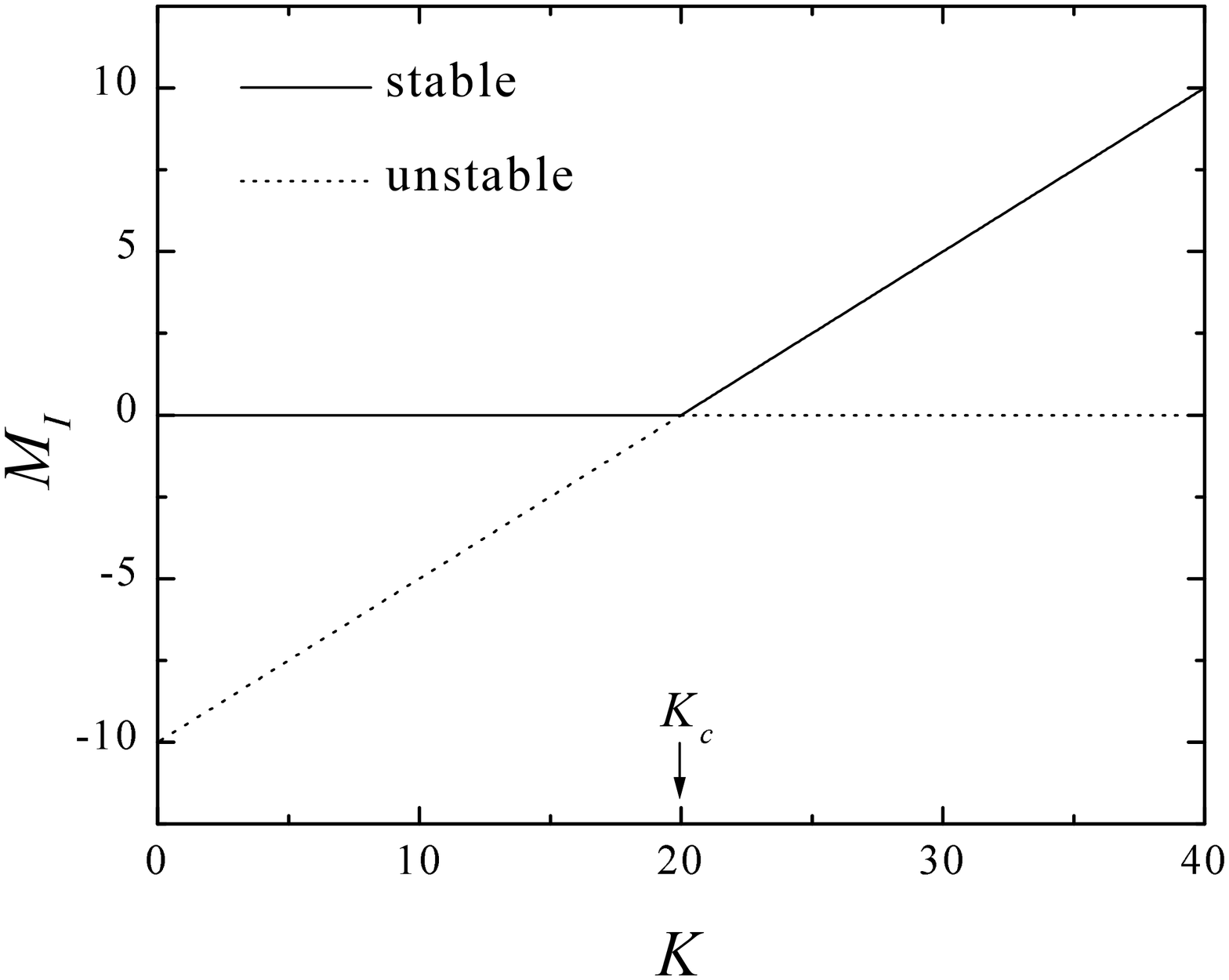}}
\caption{Bifurcation diagram of the density of infected mice $M_I$, as a function of the
carrying capacity $K$. Model parameters are: $a=0.1$, $b=1$, $c=0.5$.}
\label{bif}
\end{figure}

The system~(\ref{dmsdt},\ref{dmidt}) has four equilibria. Two of them are
irrelevant to the present analysis (the null state, which is always unstable,
and a state with $M_I<0$ for any parameters). The other two equilibria
interchange their stability character at a critical value of the carrying
capacity, a result that we show in Fig.~\ref{bif} as a bifurcation diagram. The
critical value of the carrying capacity is
\begin{equation}
K_c=\frac{1}{a}\left(\frac{b}{b-c}\right).
\label{kc}
\end{equation}

We can see that the prevalence of the infection can be correlated, through $K$,
with the diversity of habitats and other ecological conditions. Thus, a
scarcity of resources---that is to say, a low value of $K$---is accompanied by
a lower number of infected mice, as found in field studies such
as~\cite{mills99b,kuenzi99,abbot99}. Moreover, for values of $K$ below the
threshold $K_c$ the number of infected animals is effectively zero, a fact that
has also been observed in the field (see for
example~\cite{calisher99,parmenter99,mills99b}). That is, if temporarily the
ecological conditions at a place in the landscape get adverse for the mice
(because of a drought, for example) the infection can drop to zero.
Correspondingly, when conditions improve again the infection reappears. The
density of infected mice can even display a dramatic increase with respect to
previous years, if a rare climatic event such as El Ni\~{n}o Southern Oscillation
brings enhanced precipitation and the consequent increase in edible resources
for the mice. An El Ni\~{n}o event in 1991-1992, precisely, preceded the outbreak
of HPS in 1993 in the Southwest~\cite{glass00}.

\begin{figure}
\centering
\resizebox{\columnwidth}{!}{\includegraphics{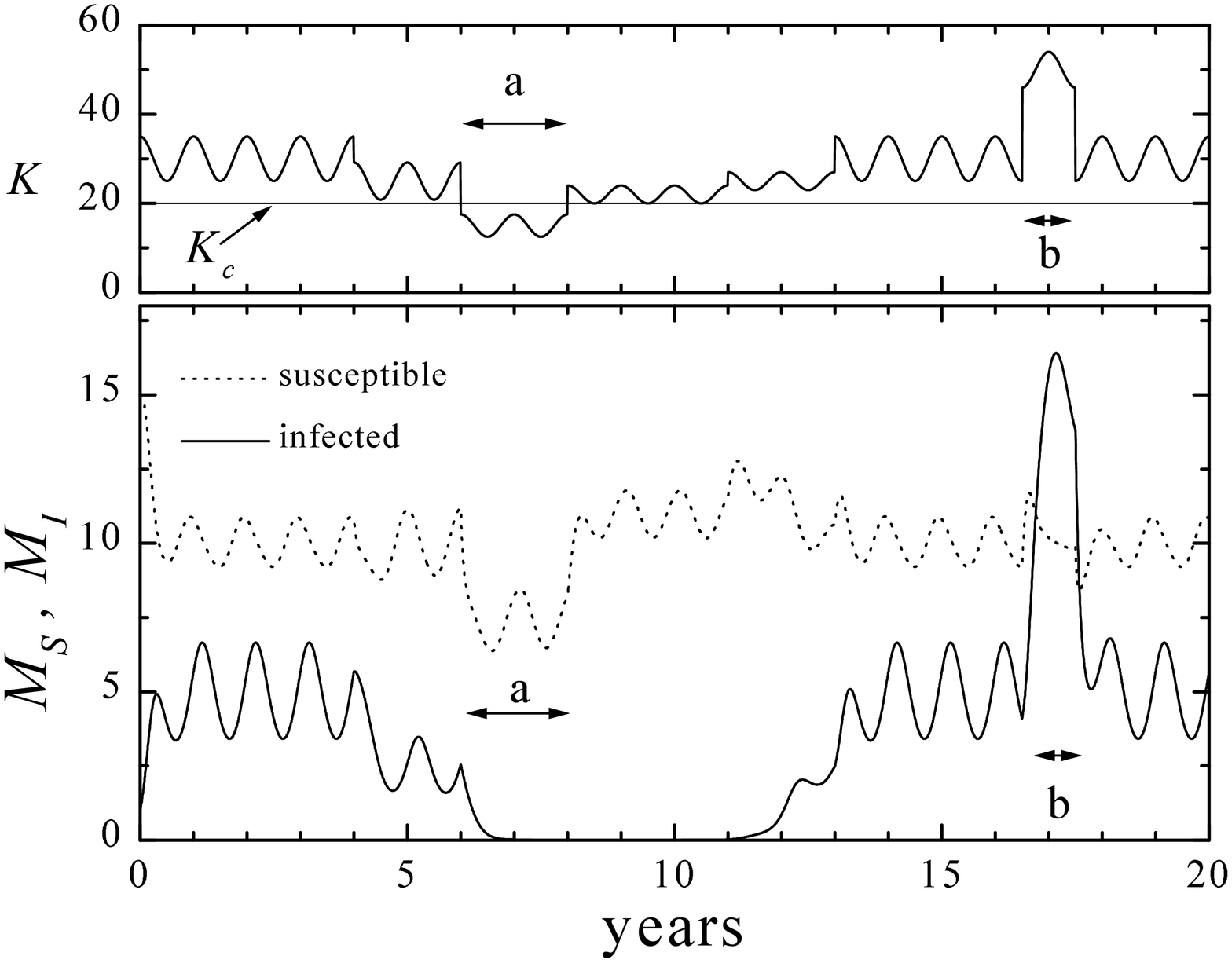}}
\caption{Temporal evolution of the population of mice (bottom) in a
caricature time-dependent carrying capacity (top). Two special events are
marked: (a) The carrying capacity is below the $K_c$ threshold (shown as a
horizontal line). (b) An extraordinary one-year event of greater carrying
capacity. Same parameters as in Fig.~\ref{bif}.}
\label{mice}
\end{figure}

Figure~\ref{mice} shows a simulation of such events, within the context of the
present model. A time-dependent carrying capacity is shown in Fig.~\ref{mice}
(top), and the corresponding values of the susceptible and infected mice
populations, $M_S(t)$ and $M_I(t)$ respectively, are displayed in
Fig.~\ref{mice} (bottom). We model the carrying capacity with a yearly
sinusoidal behavior to emulate seasonal variations. A period of 20 years is
shown, during which the carrying capacity oscillates around a value, sometimes
above $K_c$ (shown as a horizontal line), sometimes below it. Discontinuities
in the carrying capacity, some of which are present in Fig.~\ref{mice} (top),
do not necessarily occur in nature, and appear here because we keep the
modeling of $K(t)$ at an elementary level, to illustrate the main features of
the system. The period marked ``a'' in Fig.~\ref{mice} (from years 6 to 8) is
characterized by values of $K$ below $K_c$, and corresponds to very adverse
environmental conditions. During these ``bad years'' the infection level
effectively drops to zero, while the population of healthy mice, even if
reduced, subsists. A return to ``normal'' carrying capacities after year 8
produces a very slow recovery of the infected population, which attains again
appreciable values after year 11. An extraordinary event on year 17 is marked
as ``b'' in Fig.~\ref{mice}. It corresponds to an increase in the carrying
capacity (top), perhaps following an event such as El Ni\~{n}o the year before.
These improved environmental conditions are followed by an immediate (if
moderate) increase in the population of susceptible mice (bottom, dotted line),
and by a slightly delayed outbreak of infection (bottom, full line). An event
such as this would appreciably increase the risk for the human population to
become infected.

\section{Spatially extended model}

\label{extended}

The range of the deer mice is wide, comprising a diverse landscape with a
variety of habitats. This spatial extension and the inhomogeneous way in which
it affects local populations can be included in a spatially extended version of
the model, where $M_S$, $M_I$ and $K$ become functions of a space variable
$\mathbf{x}$. Diffusive movement of the mice provide an adequate mechanism of
transport, since mice of the genus \emph{Peromyscus} are known to hold a home
range during most of their adult life, occasionally shifting it to nearby
locations, in particular if these are vacant~\cite{stickel68,vessey87}. In
principle, different diffusion coefficients should be used for susceptible and
infected mice. The observation that juvenile animals are the most mobile
\cite{calisher99} and that the infection affects mainly adult males
\cite{mills99a} certainly supports this. We will choose later, however, for the
sake of simplicity of the model, to keep both diffusivities equal. The extended
model can be written as:
\begin{eqnarray}
\frac{\partial M_S}{\partial t}&=&f(M_S,M_I)+D_S\nabla^2
M_S, \label{dmsxdt}\\
\frac{\partial M_I}{\partial t}&=&g(M_S,M_I)+D_I\nabla^2
M_I, \label{dmixdt}
\end{eqnarray}
where $f$ and $g$ are the r.h.s. of Eqs.~(\ref{dmsdt}) and~(\ref{dmidt})
respectively (and contain the specific form of the spatial dependence
$K(\mathbf{x})$), and we include separate diffusion coefficients $D_S$ and
$D_I$ for the two classes of mice.

The solution of the system (\ref{dmsxdt},\ref{dmixdt}), and even its stationary
solution, may impossible to find, analytically, for an arbitrary function
$K(\mathbf{x})$. We describe below some general considerations about stability,
followed by numerical solution for $\mathbf{x}$-dependent $K$.

\subsection{Stability of the extended solutions}

\label{homogeneous}

Suppose that $M_S^*(\mathbf{x})$ and $M_I^*(\mathbf{x})$ are stationary
solutions of Eqs.~(\ref{dmsxdt},\ref{dmixdt}), i.e. they are solutions of a
Laplace equation with nonlinear, space-dependent sources:
\begin{eqnarray}
\nabla^2 M_S&=&-f(M_S,M_I)/D_S, \\
\nabla^2 M_I&=&-g(M_S,M_I)/D_I,
\end{eqnarray}
found by setting the time derivative of Eqs.~(\ref{dmsxdt},\ref{dmixdt}) equal
to zero. A perturbation around this equilibrium can be written as:
\begin{eqnarray}
M_S(\mathbf{x},t)&=&M_S^*(\mathbf{x})+u_S(\mathbf{x},t), \\
M_I(\mathbf{x},t)&=&M_I^*(\mathbf{x})+u_I(\mathbf{x},t).
\end{eqnarray}
When the two-component vector  $u=(u_S,u_I)$ describing the perturbation is
inserted into the differential equations (\ref{dmsxdt},\ref{dmixdt}), a
linearization around the equilibrium solutions yields
\begin{equation}
\frac{\partial u(\mathbf{x},t)}{\partial t}=A(\mathbf{x})\,u(\mathbf{x},t) + D\nabla^2u(\mathbf{x},t),
\label{perturb}
\end{equation}
where $A(\mathbf{x})$ is the linearization of the nonlinear terms of
Eqs.~(\ref{dmsxdt},\ref{dmixdt}) around the equilibrium, viz.,
\begin{equation}
A(\mathbf{x})=\left[
  \begin{array}{cc}
    \frac{\partial f}{\partial M_S} &  \frac{\partial f}{\partial M_I} \\
    \frac{\partial g}{\partial M_S} &  \frac{\partial g}{\partial M_I}
  \end{array}
  \right]_{\{M_S^*,M_I^*\}},
\end{equation}
and $D$ is the $2\times 2$ diagonal matrix of the diffusivities.

Solutions of Eq.~(\ref{perturb}) can be looked for in the form of plane waves,
\begin{equation}
u(\mathbf{x},t)\sim e^{i\,\mathbf{k}\cdot\mathbf{x}+\lambda t},
\end{equation}
which, in Eq.~(\ref{perturb}), satisfies:
\begin{equation}
[\lambda I -A(\mathbf{x})+k^2 D]u(\mathbf{x},t)=0,
\label{perturb2}
\end{equation}
where $I$ is the identity matrix. The nontrivial solutions of
Eq.~\ref{perturb2}) will provide a dispersion relation $\lambda(k^2)$,
implicitly:
\begin{equation}
\det[\lambda I -A(\mathbf{x})+k^2 D]=0.
\label{dispersion}
\end{equation}

\begin{figure}
\centering
\resizebox{\columnwidth}{!}{\includegraphics{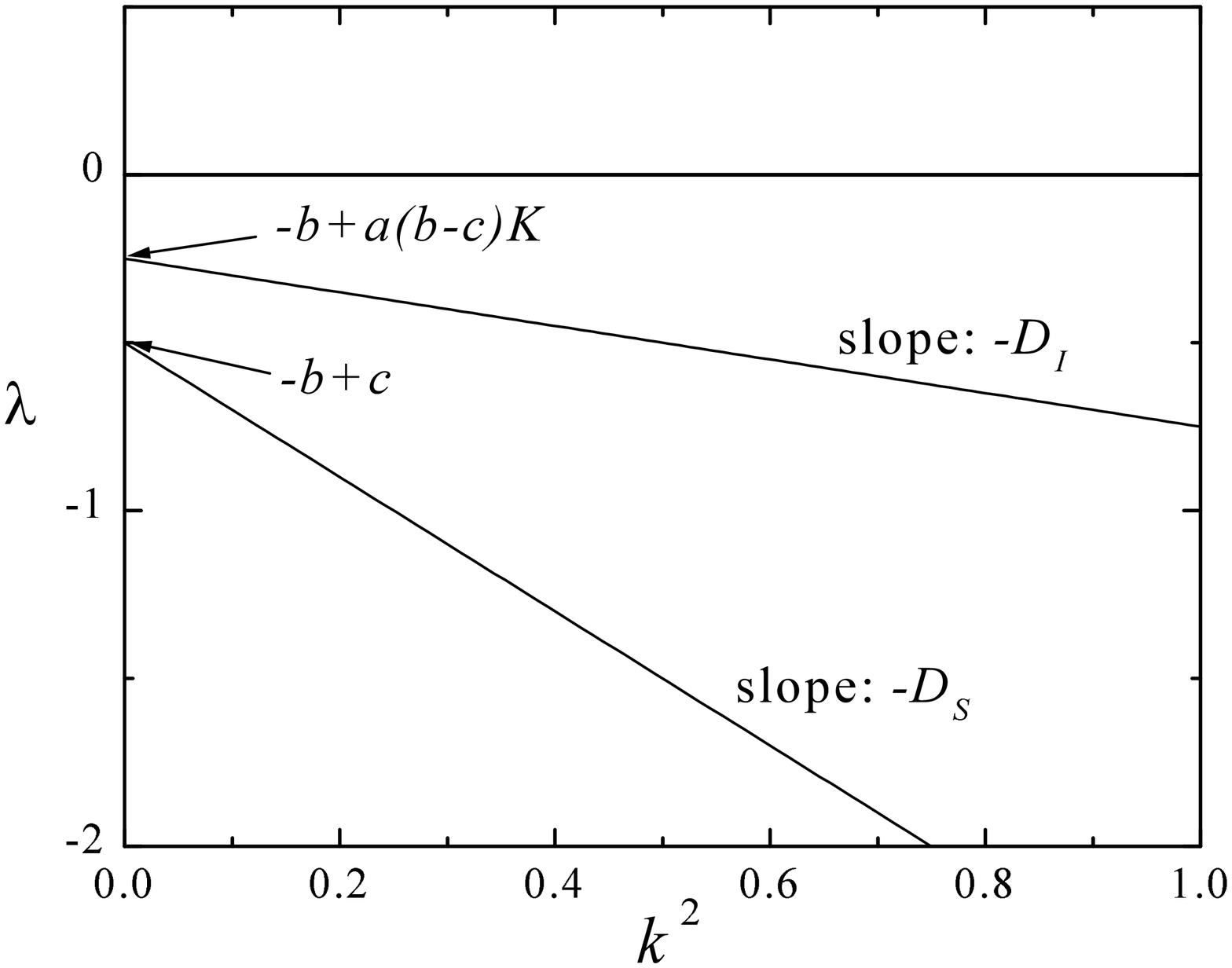}}
\caption{Dispersion relations between the temporal eigenvalue $\lambda$ and the
squared modulus of the wave number of the perturbation, $k^2$, for the two
homogeneous steady states. Model parameters as in Fig.~\ref{bif}, $K=15$,
$D_S=1$, $D_I=0.5$.}
\label{dispersfig}
\end{figure}

In the general situation of $\mathbf{x}$-dependent $K$, it is not possible to
proceed further without the knowledge of the equilibria. However, in a system
where $K$ does not depend on the space variable, an analytic assessment of the
stability of the \emph{homogeneous} steady states is possible. We have again
two relevant steady states: $\{M_S^*=(b-c)K,M_I^*=0\}$ and
$\{M_S^*=b/a,M_I^*=-b/a+(b-c)K\}$. The dispersion relations corresponding to
each one of these are easily found from Eq.~(\ref{dispersion}). Those
corresponding to the first one (the equilibrium with $M_I^*=0$) are shown in
Fig.~\ref{dispersfig}. They provide a direct stability criterion. The slopes of
the two lines are determined by the diffusion coefficients only, and as such
are always negative. It can be seen that one of the temporal eigenvalues is
always negative, provided that $b>c$, which is, obviously, the sensible case in
the biological context since otherwise no positive solutions are found. The
other eigenvalue is negative provided that $K<K_c$, which is the same stability
condition found in the nonextended case. Furthermore, when the state becomes
unstable, the fastest growing mode of the perturbation (the one with larger
$\lambda$) is that with $k^2=0$, an homogeneous perturbation. Under such
conditions, the perturbation eventually drives the system to the other
homogeneous steady state, having a nonzero infected population. In this simple
model, hence, there are no spatially dependent instabilities to the homogeneous
steady state.

\subsection{Refugia}

\label{refugia}

Certainly, the most interesting situations arise when $K$ exhibits a spatial
dependency. This is in fact the case in the field, where $K$ follows the
diversity of the landscape. We have analyzed two cases of this situation, by
means of a numerical solution of Eqs.~(\ref{dmsxdt},\ref{dmixdt}). The first
case is a 1-dimensional system, where the profile displayed by the stationary
solutions of the populations is readily accessible. The second one is a
2-dimensional system, intended to provide a more realistic picture of the
consequences of the bifurcation.

\begin{figure}
\centering
\resizebox{\columnwidth}{!}{\includegraphics{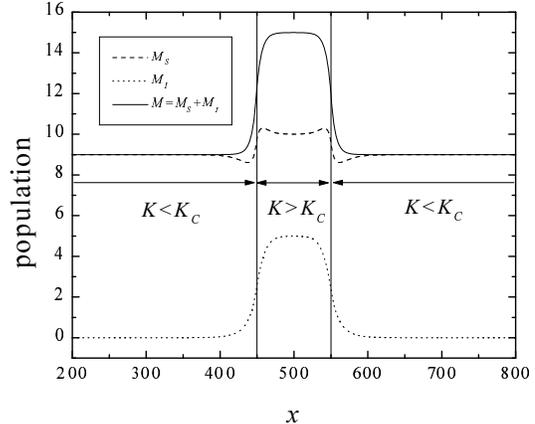}}
\caption{Stationary solution of the extended model in one dimension. The carrying
capacity $K$ consists of a spot of high value, $K>K_c$, immersed in a region of
lower capacity, $K<K_C$. The boundaries are shown as vertical lines. Model
parameters as in Fig.~\ref{bif}, $D=20$, $K=1.5K_c$ in the refugium, $K=0.9K_c$
outside of it.}
\label{refugia1d}
\end{figure}

We consider first a 1-dimensional landscape, consisting of a spot of high
carrying capacity ($K>K_c$) in the middle of a bigger region of low carrying
capacity ($K<K_c$). A typical situation is shown in Fig.~\ref{refugia1d}, where
vertical lines represent the boundaries between the three zones. From an
arbitrary initial condition of the populations, a steady state is attained in
which the infected population is concentrated at the spot of higher $K$, that
constitutes a ``refugium.'' A ``leak'' of infection is seen outside the
high-$K$ region, due to the diffusion. Far from this, the mouse population
remains effectively not infected.

\begin{figure}
\centering
\resizebox{\columnwidth}{!}{\includegraphics{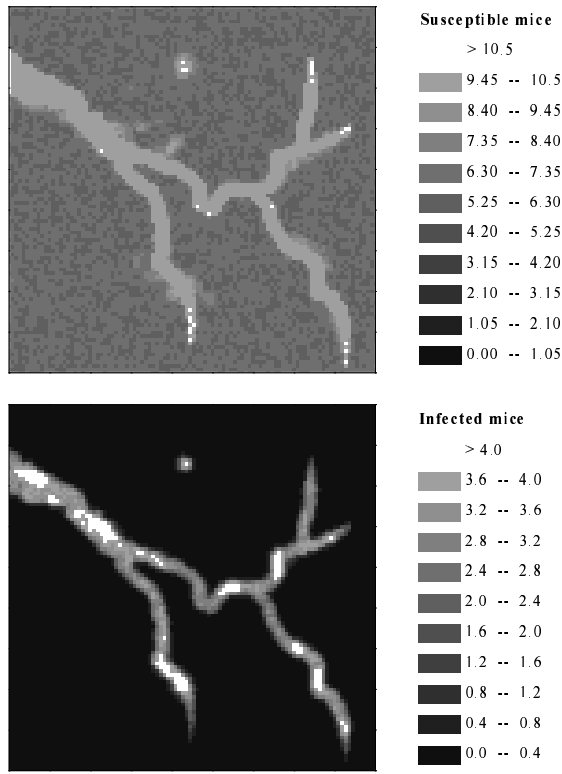}}
\caption{Stationary solution of the extended model in two dimensions. The carrying
capacity $K$ simulates a landscape where it is higher near a ``river.'' Model
parameters as in Fig.~\ref{bif}, $D=1$.}
\label{refugia2d}
\end{figure}

In Fig.~\ref{refugia2d} we show the steady state of a 2-dimensional realization
of the system~(\ref{dmsxdt},\ref{dmixdt}) on a square grid which simulates a
hypothetical landscape by assigning different values to $K_{ij}$, the carrying
capacity at each site. This is supposed higher along a ``river'' as can be
inferred from the density plots shown. The non-infected population occupies the
whole landscape, with a non-homogeneous density. Moreover, as expected from the
results of the homogeneous model, for small and moderate values of the
diffusion coefficient, the infected population survives in a patchy pattern,
only in the regions of high carrying capacity, becoming extinct in the rest.
These ``islands'' of infection become reservoirs of the virus~\cite{kuenzi99}
or ``refugia''~\cite{yates2001}, which are the places of highest risk for human
exposure and contagion of the virus. It is also from these refugia that the
disease would spread (blurring the patchiness, as observed
in~\cite{mills99b,abbot99}) when environmental conditions change. While our
model is qualitative at this stage, this is precisely what is observed in the
field. We comment in passing that the steady state distribution of neither
infected  nor susceptible mice reproduces exactly the distribution of the
carrying capacity. This is the result of the interaction of diffusion with the
nonlinear interactions. Thus, notice in the  1-dimensional representation shown
in  Fig.~\ref{refugia1d} that, although the carrying capacity follows a step
distribution, the mice populations are not steps. Both $M_S$ and  $M_I$ have
diffusive ``leaking'' , the former exhibiting a dip as one moves out of the
region of large capacity. Similarly, in the 2-dimensional case shown in
Fig.~\ref{refugia2d}, we see that the peaks of the populations represented by
pure white appear at different places for the susceptible and infected. They do
not occupy the entire ``river'' region or follow precisely the peaks of the
distribution of the carrying capacity.

\section{Concluding remarks}

Two observed characteristics of Hantavirus infection have served as the focus
of our present investigation: temporal patterns in the evolution of the
population of infected mice, and emergence of spatial features in the landscape
of infection, the so-called ``refugia.'' Our theoretical model, represented by
(\ref{dmsxdt},\ref{dmixdt}), incorporates nonlinear terms describing infection
transfer between mice populations, a logistic description of their interactions
with the environment, and diffusive terms representing their motion over the
terrain. We have shown that the combination of these various terms, while
simple, naturally predicts the temporal and spatial patterns whose observations
have motivated the analysis. Our tools of investigation comprise of analytic
stability considerations which result in features such as bifurcation behavior
(e.g., Fig.~\ref{bif}) as well as numerical procedures which yield the temporal
evolution (e.g., Fig.~\ref{mice}). The spatial extension inherent in our model
allows us to analyze the dispersion relation describing in a simplified case
departures from stationary states (see Fig.~\ref{dispersfig}) and to deduce
more generally the existence of the ``refugia'' (see
Figs.~\ref{refugia1d},~\ref{refugia2d}).

We are currently in the process of investigating a number of further features
of the spread of infection on the basis of the model and techniques explained
in the present paper. They include among others: traveling waves which can
depict the spread of fronts of infection emanating from the refugia in periods
favorable to the propagation of the infection; situations in which the mice are
limited in their meanderings to more or less localized regions for territorial
reasons but spread the infection when the localized regions overlap;
non-diffusive effects in the motion of the mice over the terrain; the effect of
stochastic disturbances in the environment; and relevant details of the
infection process such as delay effects related to finite incubation periods.
The results of these investigations will be reported elsewhere.

\begin{acknowledgments}
We acknowledge many discussions with Terry Yates, Bob Parmenter, Fred Koster
and Jorge Salazar from which we learnt much regarding the peculiarities of the
hantavirus including the observation of refugia. We also thank Greg Glass, Karl
Johnson  and Luca Giuggioli for discusions. V. M. K. acknowledges a contract
from the Los Alamos National Laboratory to the University of New Mexico and a
grant from the National Science Foundation's Division of Materials Research
(DMR0097204). G. A. thanks the support of the Consortium of the Americas for
Interdisciplinary Science and the hospitality of the University of New Mexico.
\end{acknowledgments}


\begin{thebibliography}{99}
\bibitem{schmaljohn97} C. Schmaljohn and B. Hjelle, Emerging Infectious
Diseases \textbf{3}, 95 (1997).

\bibitem{mills99a} J. N. Mills, T. L. Yates, T. G. Ksiazek, C. J. Peters and J.
E. Childs, Emerging Infectious Diseases \textbf{5}, 95 (1999).

\bibitem{mills99b} J. N. Mills, T. G. Ksiazek, C. J. Peters and J. E. Childs,
Emerging Infectious Diseases \textbf{5}, 135 (1999).

\bibitem{calisher99} C. H. Calisher, W. Sweeney, J. N. Mills and B. J. Beaty,
Emerging Infectious Diseases \textbf{5}, 126 (1999).

\bibitem{parmenter99} C. A. Parmenter, T. L. Yates, R. R. Parmenter and J. L.
Dunnum, Emerging Infectious Diseases \textbf{5}, 118 (1999).

\bibitem{kuenzi99} A. J. Kuenzi, M. L. Morrison, D. E. Swann, P. C. Hardy and
G. T. Downard, Emerging Infectious Diseases \textbf{5}, 113 (1999).

\bibitem{yates2001} Terry L. Yates, personal communication (2001).

\bibitem{anderson} R. M. Anderson and R. M. May, \emph{Infectious diseases of
humans, Dynamics and control} (Oxford University Press, Oxford, 1992).

\bibitem{murray} J. D. Murray, \emph{Mathematical Biology}, 2nd ed. (Springer,
New York, 1993).

\bibitem{terman68} C. R. Terman, in \emph{Biology of \emph{Peromyscus}
(Rodentia)}, J. A. King (editor) (The American Society of Mammalogists, Special
publication No. 2, 1968).

\bibitem{abbot99} K. D. Abbot, T. G. Ksiazek and J. N. Mills, Emerging
Infectious Diseases \textbf{5}, 102 (1999).

\bibitem{glass00} G. E. Glass \emph{et al.}, Emerging Infectious Diseases
\textbf{6}, 238 (2000).

\bibitem{stickel68} L. F. Stickel, in \emph{Biology of \emph{Peromyscus}
(Rodentia)}, J. A. King (editor) (The American Society of Mammalogists, Special
publication No. 2, 1968).

\bibitem{vessey87} S. H. Vessey, American Zoologist \textbf{27}, 879 (1987).

\end{thebibliography}
\end{document}